\documentclass{Interspeech}

\interspeechcameraready 

\title{Relationship between objective and subjective perceptual measures of speech in individuals with head and neck cancer}

\author[affiliation={1}]{Bence Mark}{Halpern}
\author[affiliation={2,3}]{Thomas}{Tienkamp}
\author[affiliation={4,5}]{Teja}{Rebernik}
\author[affiliation={6}]{Rob J.J.H.}{van Son}
\author[affiliation={2}]{Martijn}{Wieling}
\author[affiliation={2}]{Defne}{Abur}
\author[affiliation={1}]{Tomoki}{Toda}

\affiliation{}{Nagoya University}{Japan} 
\affiliation{CLCG}{University of Groningen}{the Netherlands}
\affiliation{}{University Medical Center Groningen}{the Netherlands}
\affiliation{LPP}{CNRS/Sorbonne Nouvelle}{France}
\affiliation{BCLS}{Vrije Universiteit Brussel}{Belgium}
\affiliation{Department of HNO}{Netherlands Cancer Institute}{the Netherlands}
\email{halpern.bence.e8@f.mail.nagoya-u.ac.jp, tomoki@icts.nagoya-u.ac.jp}

\usepackage{comment}

\begin{document}

\maketitle

\begin{abstract}
%TT: Max 1000 characters

Meaningful speech assessment is vital in clinical phonetics and therapy monitoring. This study examined the link between perceptual speech assessments and objective acoustic measures in a large head and neck cancer (HNC) dataset. Trained listeners provided ratings of intelligibility, articulation, voice quality, phonation, speech rate, nasality, and background noise on speech. Strong correlations were found between subjective intelligibility, articulation, and voice quality, likely due to a shared underlying cause of speech symptoms in our speaker population. Objective measures of intelligibility and speech rate aligned with their subjective counterpart. Our results suggest that a single intelligibility measure may be sufficient for the clinical monitoring of speakers treated for HNC using concomitant chemoradiation.

\end{abstract}

\section{Introduction}

Meaningful assessment of speech through acoustic measures is important both in medical decision making and in clinical phonetics. In the decision making, the measures obtained through speech assessment directly influences speech therapy monitoring and planning. In clinical phonetics, these measures are crucial for reproducible research.

Assessing speech, whether in clinical or research settings, typically involves two types of measures: subjective (perceptual) evaluation and objective (computational) evaluation. On the one hand, subjective evaluations rely on trained listeners rating various aspects of speech, for example, intelligibility (here: the degree to which speech is understood) or phonation (here: accurate voicing distinction). However, these evaluations can be time-consuming, require trained raters, and may be influenced by biases such as listener familiarity \cite{landa2014association} and professional experience \cite{de1997test}. On the other hand, objective evaluations use computational methods and algorithms to analyse speech signals and derive quantitative measures. These methods offer the potential for automated, consistent, and rapid assessment. However, a persistent challenge in objective evaluation, besides the lack of interpretability of most methods, is ensuring that the chosen metrics mimic human perception and are clinically relevant.

%A common practice in developing objective speech measures is to focus on a single perceptual dimension, such as articulatory clarity, and validate it by demonstrating a strong correlation with perceptual ratings \cite{janbakhshi2019pathological, middag2009automated}. % CAMERA READY: PUT BACK XPPG-PCA REFERENCE
%However, high correlation does not guarantee that the measure specifically captures the percept. Perceptual dimensions corresponding to different parts of speech productions system, e.g. articulatory clarity and voice quality, are often correlated due to a common underlying factor, such as overall disorder severity \cite{tu2016relationship}. These interdependencies between perceptual dimensions can lead to correlations between their respective ratings, even though voicing and articulation problems are attributed to distinct parts of the speech production apparatus. This is the classic "correlation does not imply causation" problem.

% CAMERA READY: PUT BACK XPPG-PCA REFERENCE
A common practice in developing objective speech measures is to focus on a single perceptual dimension, such as intelligibility. The objective speech measure is then validated by showing a strong correlation between this measure and subjective perceptual ratings \cite{janbakhshi2019pathological, middag2009automated}. The goal is often to develop measures specific to either the articulatory subsystem (e.g., articulation) or the laryngeal subsystem (e.g., phonation, voice quality). The articulatory subsystem contains the structures and processes that shape the vocal tract for speech (e.g., tongue, lips, jaw), while the laryngeal subsystem contains the structures and processes that generate the voice source (e.g., vocal fold vibration). However, a high correlation between an objective speech measure and a perceptual rating does not guarantee that the measure specifically captures the intended percept. Perceptual dimensions corresponding to these different subsystems, for example, articulatory clarity and voice quality, are often correlated due to a common underlying factor, such as overall speech disorder severity \cite{tu2016relationship}. These interdependencies between perceptual dimensions can lead to correlations between their respective ratings, even though voicing and articulation problems are attributed to distinct parts of the speech production apparatus. Furthermore, other factors not related to the produced speech can also play a role. For example, noise \cite{eadie2021effect} has also been shown to have an impact on intelligibility ratings. 

To this end, our present study investigates a range of perceptual measures across the articulatory and laryngeal subsystems: intelligibility, articulation (clarity of pronunciation), voice quality (overall vocal characteristics), phonation, rate (rate of speech), nasality (resonance in the nasal cavity), and the presence of background noise (extraneous sounds).
Specifically, we investigated whether these perceptual measures correlate with objective measures and each other on longitudinal audio recordings from 53 Dutch individuals with head and neck cancer (HNC). This represents 2\% of the annual Dutch HNC population \cite{iknl2023kankerregistratie}. We investigated the following research questions: 

\begin{enumerate}[label=\textbf{RQ\arabic*},noitemsep]
\item What is the correlation between commonly used perceptual measures of speech with each other?
\item How well can these subjective evaluation measures be predicted by objective evaluation measures? (as measured by Pearson’s correlation)
%\item Do objective measures show the same interrelatedness between perceptual domains as subjective measures?
\end{enumerate}

\section{Related works}
\label{section:related}

This section reviews previous studies that investigated the relationships between multiple perceptual measures, as well as those that focused on the impact of one perceptual measure to another. However, note that most of these are studies on non-HNC populations. Due to limited space, we are not able to give a full account of the objective evaluation measures, and the reader is referred to the review of \cite{halpern2023automatic}. 

\subsection{Interrelationships between perceptual measures}
Tu and colleagues conducted a study comparing a number of perceptual measures (articulatory precision, nasality, vocal quality, severity, and prosody) \cite{tu2016relationship}. Their study included 32 speakers with dysarthria rated by 15 second-year speech-language therapy master students. The lowest correlation was between vocal quality and nasality ($r=0.69$), the highest correlation was between vocal quality, articulation precision, and severity (all of them $r=0.91$), showing strong interrelatedness between measures overall. In another study, the De Bodt et al. \cite{de2002intelligibility} showed a strong correlation between articulatory precision and intelligibility ($r = 0.82$), and a moderate correlation between voice quality and intelligibility ($r=0.46$), and weak correlation between nasality and intelligibility ($r=0.32$) in speakers with dysarthria ($n=79$).
%In another study, De Bodt et al.\cite{de2002intelligibility} argued that intelligibility values can be predicted as a linear combination of voice quality, articulation, nasality and prosody in speakers with dysarthria (n = 79). A model was put forward in the paper where articulation had the largest weight, and nasality had the lowest weight, with an $R^2=0.81$ \cite{de2002intelligibility}.

\subsection{Impact of perceptual measures on intelligibility}

\textbf{Speech rate}: %The relationship between speech rate and intelligibility has been hard to exactly measure because humans cannot talk as fast as the perception allows to understand. An obvious work around would be speeding up the speech, however this has been historically difficult. Nowadays, fairly natural ways to speed up and slow down recorded speech is available using computerised techniques. 
In the case of typical speech, faster speech is usually more difficult to understand \cite{du2014effect}. For pathological speech, the relationship is a bit different. For example, for speakers with dysarthria, an atypical (either too fast or too slow) speech rate is often used as a diagnostic criterion \cite{blanchet2009speech}. For certain speakers with dysarthria who speak too fast, teaching them to speak slowly improves intelligibility \cite{van2009effect}.

\textbf{Articulatory precision}: Articulatory precision and intelligibility have recurrently been positively associated with each other. Apart from De Bodt et al. \cite{de2002intelligibility} results mentioned above, Thompson and Kim \cite{thompson2024acoustic} also reported strong correlations ($r=0.9$) between intelligibility and articulatory precision in 40 speakers with and without dysarthria.

%Based on the linear part of the sigmoid curve, a high correlation ($r=-0.72$) can be calculated 
%The relationship is impacted by the presence of pathology but in general holds true. For example, in the case of dysarthric speech, decreasing the speech rate is actually a very effective way to improve intelligibility \cite{prananta22_interspeech}.

%Nasality is one of the few perceptual measures that have a widely accepted instrumental objective measure by the nasometer \cite{liu2022correlation}.
\textbf{Nasality}: The literature reports varying correlations between intelligibility and nasality ratings. McWilliams at al. \cite{mcwilliams1954some} found a high correlation between nasality and intelligibility ratings ($r=0.72$) of 48 cleft-palate patients rated by seven listeners. In Cantonese children with cleft palate, no significant correlation was established neither for nasal ($r=-0.38$), nor for non-nasal sentences ($r=-0.41$)\cite{chun2001relationship}.

% and in general high interrater correlation was observed (\texttt{Spearman-Brown}: $0.925$).
\textbf{Noise}: A common finding is that extraneous noise negatively affects intelligibility. Depending on the type of noise, there are some intricacies to this effect, i.e., noise with linguistic component in the same language (e.g., babble noise) is more detrimental to intelligibility compared to noise without linguistic content (e.g., white noise) \cite{lee2009effects}. Studies on speakers with dysarthria \cite{mcauliffe2009effect} or HNC \cite{eadie2021effect} have both shown that babble noise impacts pathological speech more than typical speech.

\section{Dataset and Experimental Percepts}
\label{section:dataset}
We use the NKI-SpeechRT dataset, introduced in \cite{halpern2024reference}. The dataset contains speakers with mild-to-mid severity pre and post-treatment for HNC with concomitant chemoradiotherapy (CCRT). The dataset includes 55 speakers (45 male, 10 female; mean age = 57 years, range = 32-79 years), of whom 47 are native speakers of Dutch.
The remaining 8 speakers are non-native speakers. The dataset includes recordings from a speaker at a maximum of five time points: before CCRT, ten weeks post-CCRT, and 12 months post-CCRT. The other two stages are unknown. The total number of combination of speakers and stages (e.g., \texttt{speaker1\_pre}, \texttt{speaker1\_post\_12}, from now on: speaker-stages) are 141 (pre-CCRT 54, post-CCRT 87). There are no typical speakers in the dataset. For the current data analysis, 5 speaker-stages (2 pre-CCRT, 3 post-CCRT), and two speakers (both male) are excluded due to the inability to get forced alignments. In total 136 speaker-stages, and 53 speakers are included. Speaker-stages are used for the correlation experiments. The total recorded audio is ca. 4 hours. 

% 31T8YAYX_pre, X9USMFSI_post-3, NZBFE42K_pre, YKH6GNWA_post-1
% 0SJ8H3KB_post-1

% \cite{bomans2017vijvervrow} if there is place
Participants were asked to read the Dutch text 'De vijvervrouw’ by Godfried Bomans. Recordings were made with a Sennheiser MD421 Dynamic Microphone and portable 24-bit digital wave recorder (Edirol Roland R-1). The audio samples were cut for surrounding silences during manual annotation, therefore no additional voice activity detection was applied. The speech samples were energy normalised to -10 dB, downsampled to 16 kHz and quantized to 16-bit PCM for the analysis.

\subsection{Subjective measures}
In a 70-minute online listening test, 14 Dutch recent speech language pathology graduates without any self-reported hearing difficulties rated the entire speech text cut into three segments of approximately equal lengths. The audio was presented at 70 dB using Sennheiser HD418 headphones. Each segment received a single rating from the 14 listeners, and the mean scores are used for the correlation experiments. All listeners rated all the stimuli/speakers.

Several dimensions, listed below, were rated simultaneously. The relevant experimental details here are reproduced based on \cite{clapham2012nki,clapham2014developing}. Ratings statistics and interrater correlations (\texttt{ICC2, K}) are in Table~\ref{tab:speech_stats}.

\noindent\textbf{Intelligibility (INT)}: Listeners were asked to rate the speech intelligibility on a 7-point scale (1 = completely unintelligible, 7 = good). The listeners were able to check the text with the ability to replay the stimuli. This allowed them to more accurately judge the intelligibility.

\noindent\textbf{Phonation (PHO)}: Listeners were asked to rate the degree to which phonation deviated from what they considered normal on a 5-point scale (1 = very deviant, 5 = normal). 
% To ask from Rob: This is what I found in the instrucitons but to me this contradicts with the paper a bit. "Accurate voicing onset for the production of sounds means that voice sounds are truly voiced (/b/d/a/) and voiceless sounds are genuine voieless. Accurate voicing onset menas that the speaker makes correct dictions between voiced and voiceless sounds compared to normal speech.TRa

\noindent\textbf{Articulatory precision (AP)}: Listeners were asked to evaluate the general precision of vowel and consonant production as compared to normal running speech on a 5-point scale (1 = extremely imprecise, 5 = normal/precise). Precise articulation was defined as correct manner and place of production and clear coordination between sounds. 

%\noindent\textbf{Accentedness (ACT)}: Listeners were asked to evaluate the weight of the speaker’s dialect or accent as compared to standard Dutch (defined as the speech commonly heard on radio and television on a 5-point scale (1 = heavy accent, 5 = no accent).

\noindent\textbf{Perceived speed (SPEED)}: Listeners were asked to evaluate the speech rate on a 9-point scale (1 = slow, 5 = normal, 9 = fast).

\noindent\textbf{Voice quality (VQ)}: Listeners rated the overall impression of voice quality on a 5-point scale (1 = severely deviated, 5 = normal). Listeners were explicitly asked to not rate pleasantness but rather the degree of voice deviation compared to normal voice.

\noindent\textbf{Nasality (NAS)}: Listeners were asked to evaluate nasality on a 5-point scale (1 = very nasal, 5 = normal). 

%\subsection{Additional noise experiment (NOISE)}
\noindent\textbf{NOISE:} A separate study was done with one expert phonetician (R.v.S.) who rated the noisiness of the recordings on a 3-point scale. Zero meant no or barely any audible noise, one meant audible noise, and two meant noisy, including sometimes other voices or ringing of the telephone. We decided to keep all the recordings even those rated very noisy for the further experimentations.

\begin{table*}[h]
    \centering
\resizebox{\textwidth}{!}{%
    \begin{tabular}{lccccccc}
        \hline
        & NAS (1-5) & PHO (1-5) & SPEED (1-9) & AP (1-5) & INT (1-7) & VQ (1-5) & NOISE (0-2) \\
        \hline
        Mean $\pm$ Std & $4.42 \pm 0.29$ & $3.89 \pm 0.58$ & $5.14 \pm 0.80$ & $3.97 \pm 0.66$ & $5.51 \pm 0.97$ & $4.32 \pm 0.42$ & $0.44 \pm 0.46$ \\
        Range [Min, Max] & $[2.63, 4.77]$ & $[1.44, 4.69]$ & $[3.36, 7.48]$ & $[2.11, 4.86]$ & $[2.31, 6.73]$ & $[2.93, 4.85]$ & $[0.00, 1.67]$ \\
IQR & $[4.33,4.58]$ & $[3.67,4.24]$ & $[4.74,5.63]$ & $[3.66,4.44]$ & $[5.07,6.19]$ & $[4.19, 4.58]$ & $[0,0.67]$ \\
\hline
ICC2,K & $0.58$ & $0.91$ & $0.90$ & $0.91$ & $0.92$ & $0.78$ & N/A \\
\hline

\end{tabular}
}
   \caption{Summary statistics for different speech attributes. Note that the statistics are calculated on the averaged ratings, hence the decimals in min/max. IQR = interquartile range, ICC2,k = intra-class correlation. NAS = nasality, PHO = phonation, SPEED = speech rate, AP = articulatory precision, INT = intelligibility, VQ = voice quality, NOISE = recording noisiness.}
    \label{tab:speech_stats}
\end{table*}

%\begin{table*}[h]
%    \centering
%\resizebox{\textwidth}{!}{%
%    \begin{tabular}{lcccccccc}
%        \hline
%        & NAS & PHO & SPEED & AP & INT & ACT & VQ & NOISE \\
%        \hline
%        Mean $\pm$ Std & $4.42 \pm 0.29$ & $3.89 \pm 0.58$ & $5.14 \pm 0.80$ & $3.97 \pm 0.66$ & $5.51 \pm 0.97$ & $3.84 \pm 0.86$ & $4.32 \pm 0.42$ & $0.44 \pm 0.46$ \\
%        Range [Min, Max] & $[2.63, 4.77]$ & $[1.44, 4.69]$ & $[3.36, 7.48]$ & $[2.11, 4.86]$ & $[2.31, 6.73]$ & $[1.44, 4.86]$ & $[2.93, 4.85]$ & $[0.00, 1.67]$ \\
%IQR & $[4.33,4.58]$ & $[3.67,4.24]$ & $[4.74,5.63]$ & $[3.66,4.44]$ & $[5.07,6.19]$ & $[3.63,4.44]$ & $[4.19, 4.58]$ & $[0,0.67]$ \\
%\hline
%ICC2,K & $0.5794$ & $0.9064$ & $0.8992$ & $0.9110$ & $0.9174$ & $0.9246$ & $0.7781$ & N/A \\
%\hline

%\end{tabular}
%}
%    \caption{Summary statistics for different speech attributes. Note that the statistics are calculated on the averaged ratings, thus the decimals in min/max. IQR = interquartile range, ICC2,k = intra-class correlation}
%    \label{tab:speech_stats}
%\end{table*}

\section{Objective measures}

In this section, we introduce the objective methods that we compare to the perceptual measures. We categorised the objective measures based on what they are intended to measure. For the objective analysis, the texts were manually cut into 23 utterances. Individual ratings for the utterances were obtained with each method, and averaged to obtain a speaker-stage level score for the correlation analysis.

\label{section:measures}
\subsection{Intelligibility estimation methods}

In the intelligibility estimation methods, we aimed to compare both reference-based and reference-free methods. The phoneme error rate needs written transcription of the audio (written reference), the neural acoustic distance needs audio reference and transcriptions (written and speech reference), and the XPPG-PCA does not need any reference.

\textbf{Phoneme error rate (PER)}: To obtain prediction for the phonemes in the utterances, we used a Dutch phoneme recogniser\footnote{\url{https://huggingface.co/Clementapa/wav2vec2-base-960h-phoneme-reco-dutch}} pre-trained on the Dutch Common Voice dataset \cite{ardilacommon}. All of our datasets had word-level transcriptions, which we converted to phoneme-level transcriptions using the Dutch \texttt{espeak} frontend of \textit{phonemizer} \cite{Bernard2021}.

%\cite{halpern2024_xppgpca}. The method previously has shown good performance on this dataset. %Open source code is available\footnote{\url{https://github.com/karkirowle/xppg-pca}}.

\textbf{Neural acoustic distance (NAD):} NAD was initially proposed as a pronunciation evaluation distance measure \cite{bartelds2022neural}. As the \textit{wav2vec-large} feature used by this distance measure has shown to be sensitive to pathological speech, too \cite{cai2024voice}, we think it can work as an intelligibility measure. First,
for each of utterance, we obtained word boundaries using the pre-trained \texttt{dutch\_cv} acoustic model from the Montreal Forced Aligner (MFA) \cite{mcauliffe17_interspeech}. Then, speech features were extracted from utterances using layer 10 of the \textit{wav2vec2-large} model, as this provided the best result in \cite{bartelds2022neural}. The segmentation on the MFA was applied on the wav2vec features to extract feature sequences for individual words in an utterance. For the distance calculation, each word (from now: target word) in an utterance was systematically compared against the same word from all the other speakers in the dataset (from now: reference words). These comparisons were performed using dynamic time warping to match the naturally varying word durations. The scores from the target words were then first averaged across all reference words, then the word-level scores in an utterance. Open source code available\footnote{\url{https://github.com/Bartelds/neural-acoustic-distance}}. 

\textbf{XPPG-PCA (PCX):} XPPG-PCA is a novel method for evaluating speech severity, combining x-vectors and phonetic posteriorgrams. These features are extracted from each utterance, normalized, and concatenated.  Principal component analysis (PCA) is then applied to this combined feature set, estimated on NKI-OC-VC \cite{halpern2023improving}, to identify dominant variations related to speech severity. The first principal component is then used to calculate a reference-free score for each utterance in a new dataset, reflecting the degree of deviation from typical speech patterns \cite{halpern2024_xppgpca}. Open source code available\footnote{\url{https://github.com/karkirowle/xppg-pca}}.
\subsection{Speed estimation methods}

%% TODO: Consider some syllable per minute measure too or transcribing to syllables (I think phonemiser can do it)
\textbf{Speech rate} $\mathbf{RATE_{S}}$: To calculate speech rate, we divided the total number of words in the transcription by the duration of the recording.
% Speech is typically quantified in words per minute (WPM) or syllables per minute (SPM). 

\textbf{Articulation rate} $\mathbf{RATE_{A}}$: Compared to speech rate, articulation rate excludes pauses. To estimate the total duration of the speech without pauses we use an energy-based voice activity detection, and consider all speech samples less than $20$ dB under the peak as speech frames. The total duration of these speech frames is used as a proxy for the duration excluding pauses. The articulation rate is calculated as number of words in the transcription divided by the duration excluding pauses.
%\subsection{Naturalness/Speech Quality estimation methods}

%While subjective speech quality assessment is not included in the percepts, it is still interesting to see how objective speech quality scores relate to other subjective percepts.

%\textbf{SSL-MOS} is \texttt{wav2vec}-model finetuned for the evaluation of listener opinions on synthetic speech quality. The architecture of the model is published in \cite{cooper2022generalization} with minor implementation changes done by \cite{huang2024mos}. The model is available online\footnote{\url{https://github.com/unilight/sheet}}

\subsection{Noise estimation methods}

$\mathbf{SNR_{N}}$: The NIST (National Institute of Standards and Technology) signal-to-noise ratio (SNR) estimation method \cite{nist_snr_measurement} uses sequential Gaussian mixture estimation to model noise. It generates a short-time energy histogram, which is then used to determine the energy distributions of both the signal and noise, from which the SNR is calculated.

$\mathbf{SNR_{W}}$: Waveform Amplitude Distribution Analysis signal-to-noise-ratio (WADA-SNR) is a reference-free SNR estimation method. This method assumes that clean speech has a Gamma distribution while the additive noise is Gaussian \cite{kim08e_interspeech}.

\section{Results}
\label{section:results}
\begin{figure}
\includegraphics[width=\columnwidth]{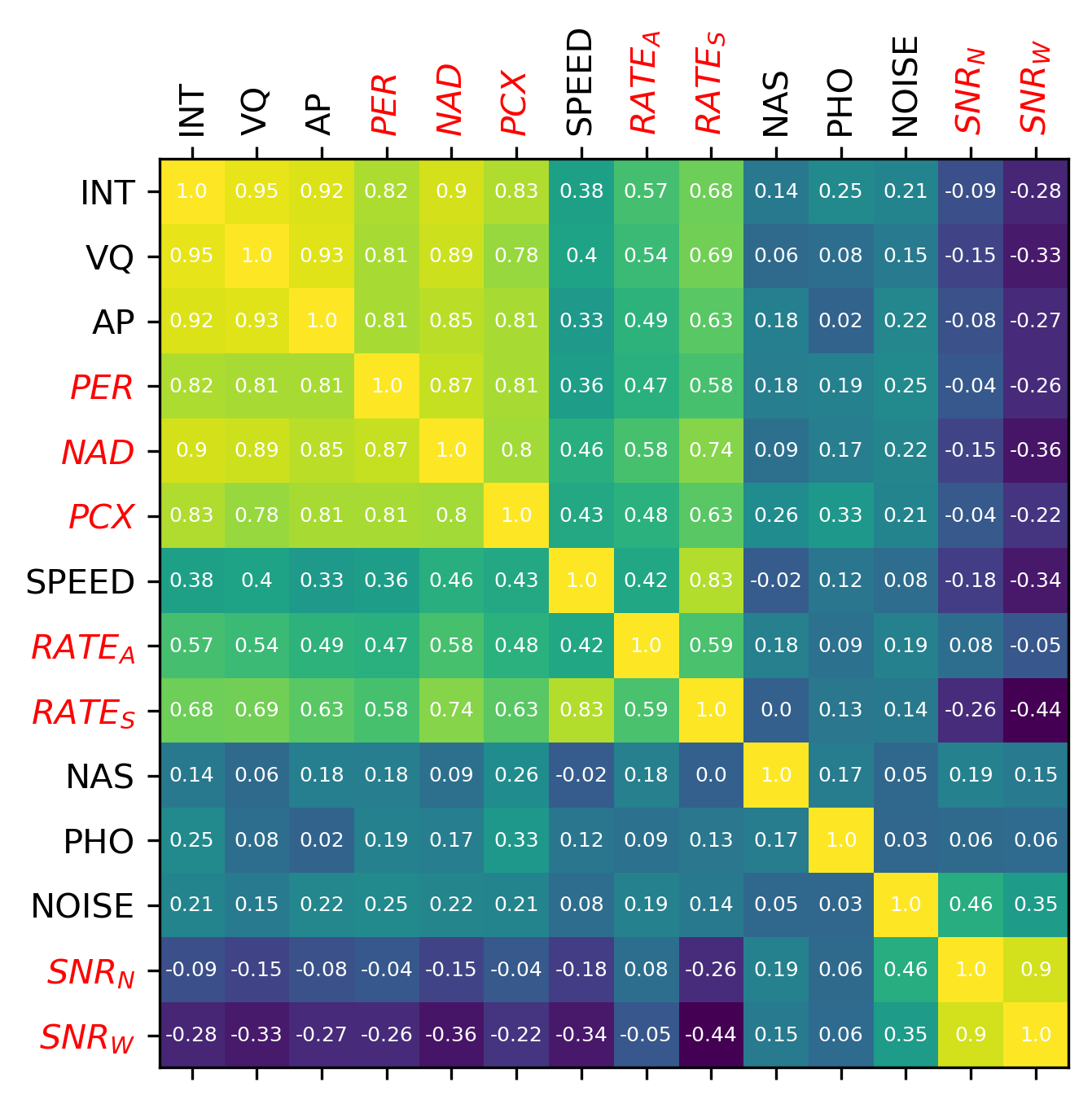}
\caption{Correlation matrix of the perceptual (subjective, in black) and objective measures (in red/italic).}
\label{fig:correlation_matrix}
\end{figure}

Please note that we have inverted the sign for the NAD, PER and the NOISE to make all measures the same directionality. 
\subsection{RQ1: Correlations between subjective measures}
As there are a large number of correlations in Figure~\ref{fig:correlation_matrix}, we will only comment on the perceptual measure correlations with intelligibility (INT), as INT showed overall the strongest correlations. There was a very strong correlation of INT with VQ ($r=0.92$) and AP ($r=0.95$). The correlation between SPEED and INT was moderate and positive ($r=0.38$), with faster speech being more understandable for the raters. Phonation had a weak correlation with intelligibility ($r=0.25$). Noise showed only a weak correlation with intelligibility ($r=0.21$).  Nasality had a none-to-weak correlation with intelligibility ($r=0.14$).

\subsection{RQ2: How well objective measures predict subjective measures?}

The objective intelligibility measures correlated well with INT. NAD achieved the best performance ($r=0.9$), followed by PCX ($r=0.83$), and finally PER ($r=0.82$). SPEED showed a strong positive correlation with $\text{RATE}_{S}$ ($r=0.83$), and a moderate positive correlation with  $\text{RATE}_{A}$ ($r=0.42$). The objective noise measures were moderately correlated with their subjective counterpart, with the $\text{SNR}_{N}$ showing a higher correlation ($r=0.46$) than $\text{SNR}_{W}$ ($r=0.35$). %We think that the $SNR_N$ is likely more suitable for scenarios where talker interference can be a noise source.

%\textbf{Speech quality}: There was no  naturalness measure in our perceptual study, so we were not sure what to expect beforehand. It turned out that naturalness measure has a moderate correlation with phonation ($r=0.50$) and noise ($r=0.43$).
%\subsection{RQ3: Behaviour of objective methods}

%In case of the intelligibility-measures, the objective measures behaved similarly as the subjective measures. In case of the SPEED measures, $RATE_{S}$ exhibited a high correlation with SPEED ($r=0.83$), while having a correlation with INT ($r=0.68$), VQ ($r=0.69$), AP ($r=0.63$). However, SPEED itself had only weak correlations with INT ($r=0.38$), VQ ($r=0.4$), AP ($r=0.33$). It is not clear what caused this discrepancy. 

\section{Discussion}

% In this work, we assessed the correlation between perceptual measures of speech with each other (RQ1), and how well these perceptual measures can be predicted by objective methods (RQ2). 
\subsection{RQ1: Correlations between subjective measures}
Regarding RQ1, our results indicated strong correlations between the measures of intelligibility (INT), articulatory precision (AP), and voice quality (VQ). 
This indicates that, within this HNC speaker population, these aspects of speech, although originating from distinct speech motor subsystems, tend to deteriorate concurrently, most likely due to radiation treatment's effect on both the articulatory \cite{jacobi2013acoustic} and laryngeal \cite{kraaijenga2016assessment} subsystems. This finding has several implications. First, for clinical practice, the strong correlation suggests that a sole intelligibility measure may be sufficient for clinical tracking in many speakers with HNC, and as shown by Tu et al. \cite{tu2016relationship}, in many speakers with dysarthria, too. Second, the strong correlations raise the risk of the 'common cause fallacy' during measure development. Developing a targeted measure, e.g., for articulation, needs validation on a population where the subjective measures are not correlated due to this common cause fallacy.

%to ensure the isolation of the intended speech attribute from other correlated aspects of speech.

The moderate correlation of SPEED and INT was partially surprising as slower speech of typical speakers is generally easier to understand compared to faster speech \cite{du2014effect}.
However, sometimes individuals with speech difficulties have trouble reaching some articulatory targets, resulting in slower speech. Therefore, it could be that speakers who are more severely affected have to slow their speech to a greater extent compared to speakers who are less severely affected \cite{tienkamp2024effect}.
%the targeted speech attribute should be isolated.

No strong correlations were found between INT and phonation (PHO), INT and  nasality (NAS), and INT and NOISE. With nasality, the poor rater agreement could be one reason why correlation could not be established. The overall typical scores on nasality with low variances may also explain why no robust correlations were observed, i.e., likely the cohort did not have nasality issues.
The phonation result also suggests that these voicing distinctions do not seem to influence intelligibility too much.  For noise, the moderate correlation is due to a single listener rating the noise.

\subsection{RQ2: How well objective do measures predict subjective measures?}
Turning to RQ2, our results show that objective measures showed strong correlations with their subjective counterpart, with the exception of noise. For intelligibility, we found that the performance depended on the reference type used. The lower correlation of PCX and INT compared to NAD and INT can be attributed to the fact that PCX does not need a reference; the lower correlation between PER and INT than between NAD and INT indicates that acoustic references are likely better than written ones. The speed measures were different in their correlation, with $\text{RATE}_{S}$ having a higher correlation. We would have expected $\text{RATE}_{A}$ to have a higher correlation than the $\text{RATE}_{S}$ due to the pauses influence on the perception but this was not the case. 
In general, the strong correlations show that objective measures are promising for clinical use, offering a potentially more consistent and less subjective way to assess speech of individuals with HNC. This holds true even considering the inclusion of non-native speakers and the presence of some noisy samples in the dataset, which shows robustness of the objective methods. However, subjective measures of nasality (NAS) and phonation (PHO) still lack reliable correlations with objective methods. With nasality, the poor rater agreement could be one reason why it is challenging to develop such measures. In contrast, phonation showed excellent inter-rater agreement yet still no correlation was found between phonation and any objective method. This suggests that an objective measure is attainable with focused research effort. 

%The low standard deviation paired with the high mean in the nasality scores suggest that the difficulty with measuring this attribute can be simply due to these speakers having mostly typical speech when it comes to nasalance. 
\subsection{Limitations and future work}
Limitations of our study include only assessing individuals with HNC, and the lack of nasality and phonation measures. We could not include the nasal severity index for nasality in the current work as it requires sustained vowels, which was not available to us \cite{bettens2016relationship}. For the phonation measure, we are not aware of any existing method specifically looking at voicing distinctions. Developing nasality and phonation measures on running speech should be part of future aims. 

Despite a great performance of the objective methods several key challenges remain. The most important is interpretability, i.e., both NAD and XPPG-PCA use neural network-based features that are not transparent enough for clinical practice. Another limitation is that all of our models are Dutch. It would be desirable to transition to language-independent models. Finally, all methods used read instead of spontaneous speech, which may be not representative of everyday speech use.

%and that we have not used an objective measure for phonation and nasality. We have not included these, as most of these need very specific stimuli, and in the current study we only used read speech.

\section{Conclusion}

Our study found strong correlation between intelligibility, voice quality, and articulation in individuals with HNC which is consistent with previous findings in speakers with dysarthria. Objective measures showed promising predictive capabilities, particularly the NAD and XPPG-PCA methods, which effectively estimated intelligibility, voice quality, and articulatory precision. Future work should focus making current  measures language independent and interpretable. 
%expanding the dataset, exploring additional objective measurement techniques, and further investigating the underlying mechanisms driving these perceptual correlations. The findings have significant implications for developing more nuanced tools in speech technology, clinical diagnostics, and interdisciplinary speech research.

\section{Acknowledgements}
The data collection in the paper received ethical approval. The Department of Head and Neck Oncology and Surgery of the Netherlands Cancer Institute receives a research grant from Atos Medical (H\"orby, Sweden). This work was further supported by the Research School of Behavioral and Cognitive Neurosciences of the University of Groningen. This work is partly financed by the NWO under project number 019.232SG.011, and 
partly supported by a project, JPNP25006, commissioned by NEDO.
%JST AIP Acceleration Research JPMJCR25U5, Japan.

\bibliographystyle{IEEEtran}
\bibliography{mybib}

\end{document}